\documentclass[sigconf, nonacm]{acmart}

\begin{document}

\title[AnalogNet: CNN Inference on Analog FPSP]{AnalogNet: Convolutional Neural Network Inference on Analog Focal Plane Sensor Processors}

%% Authors and their affiliations.
\author{Matthew Z Wong}
\authornote{Both authors contributed equally to this research while at Imperial College London.}
\email{matthew.wong@live.com.sg}
\affiliation{%
  \institution{Department of Computing, Imperial College}
}

\author{Benoit Guillard}
\authornotemark[1]
\email{benoit.guillard@epfl.ch}
\affiliation{%
  \institution{School of Computer and Communication Sciences, EPFL}
}

\author{Riku Murai}
\email{riku.murai15@imperial.ac.uk}
\affiliation{%
  \institution{Department of Computing, Imperial College}
}

\author{Sajad Saeedi}
\email{s.saeedi@ryerson.ca}
\affiliation{%
  \institution{Ryerson University}
}

\author{Paul H J Kelly}
\email{p.kelly@imperial.ac.uk}
\affiliation{%
  \institution{Department of Computing, Imperial College}
}

%% Use full list of authors in the page headers, or 
%% a more concise list ?
%\renewcommand{\shortauthors}{Wong and Guillard, et al.}

%% Abstract
\begin{abstract}
  %A key objective in computer vision is the development of high-throughput and energy-efficient vision systems. 
  We present a high-speed, energy-efficient Convolutional Neural Network (CNN) architecture utilising the capabilities of a unique class of devices known as analog Focal Plane Sensor Processors (FPSP), in which the sensor and the processor are embedded together on the same silicon chip. 
  
  Unlike traditional vision systems, where the sensor array sends collected data to a separate processor for processing, FPSPs allow data to be processed on the imaging device itself. This unique architecture enables ultra-fast image processing and high energy efficiency, at the expense of limited 
  processing resources 
  %register availability 
  and 
  %inaccurate 
  approximate computations.
  
  In this work, we 
  %introduce novel techniques 
  show how to convert standard CNNs to FPSP code, and demonstrate a method of training networks to increase their robustness to analog computation errors.
  Our proposed architecture\footnote{An implementation can be found at \url{https://github.com/brouwa/CNNs-on-FPSPs}}, coined AnalogNet, reaches a testing accuracy of \textbf{96.9\%} on the MNIST handwritten digits recognition task, at a speed of \textbf{2260~FPS}, for a cost of \textbf{0.7 mJ per frame}.

%We also experiment with two techniques to implement multi-layer CNNs on an FPSP - quantisation and pooling. The resulting accuracy is however gravely hindered by noise, for which we provide a quantitative study.
\end{abstract}

\begin{comment}

\begin{CCSXML}
<ccs2012>
<concept>
<concept_id>10010583.10010786.10010810</concept_id>
<concept_desc>Hardware~Emerging optical and photonic technologies</concept_desc>
<concept_significance>500</concept_significance>
</concept>
<concept>
<concept_id>10010147.10010178.10010224</concept_id>
<concept_desc>Computing methodologies~Computer vision</concept_desc>
<concept_significance>500</concept_significance>
</concept>
<concept>
<concept_id>10010583.10010633.10010654</concept_id>
<concept_desc>Hardware~On-chip sensors</concept_desc>
<concept_significance>300</concept_significance>
</concept>
<concept>
<concept_id>10010147.10010178.10010224.10010226</concept_id>
<concept_desc>Computing methodologies~Image and video acquisition</concept_desc>
<concept_significance>300</concept_significance>
</concept>
</ccs2012>
\end{CCSXML}

\ccsdesc[500]{Hardware~Emerging optical and photonic technologies}
\ccsdesc[500]{Computing methodologies~Computer vision}
\ccsdesc[300]{Hardware~On-chip sensors}
\ccsdesc[300]{Computing methodologies~Image and video acquisition}
\end{comment}

%% Keywords.

%% "Teaser" image. What to use here ?? The analognet_arch figure
%% is ugly on the cover...
%  \includegraphics[width=\textwidth]{figures/analognet_arch.pdf}

\maketitle

\section{Introduction}
Despite their spectacular successes, today's deep neural networks suffer from what has been termed the `inference efficiency' problem~\cite{Google_TensorflowQuantization_2018}. While these networks perform extremely well when running on specialised hardware such as GPUs, they are in many cases not fast or energy efficient enough to be effectively deployed for real-time applications on less powerful hardware \cite{NVIDIA_Acceleration_2017}. This has limited the range of potential applications, given that many artificial intelligence and (AI) robotics applications operate in real-time and yet are also power-constrained.

In this paper, we develop and implement a fast and energy efficient Convolutional Neural Network architecture on a class of devices known as Analog Focal-Plane Sensor-Processors (FPSPs). FPSPs integrate the light sensor and the early stage processing of a traditional vision system, by augmenting each photo-diode with rudimentary analog computation capabilities. Computations are said to happen \emph{on the focal plane}. This architecture reduces the need for image data to be transferred to a separate processing device, thus offering the potential  for high frame rates (1,000-100,000 FPS, frames per second) and low power consumption. On the other hand, FPSPs are very constrained in terms of memory availability, and only allow for noisy computations because of their analog nature. This paper's experimental work was conducted on the SCAMP-5 FPSP \cite{6578654}.

We introduce novel techniques to convert standard neural network layers to FPSP code, and demonstrate a method of training networks to increase their robustness to the effects of approximate computation. \textcolor{black}{To summarise, the contributions of this paper are:
\begin{itemize}
    \item a custom training process, providing neural networks parameters for an FPSP, and robust to noisy computations. The training process is general and is independent of a particular FPSP device. 
    \item an optimised and adaptable implementation of CNNs on an FPSP, relying on the strengths of the hardware to demonstrate high throughput and energy efficiency%an estimated 80\% reduction in inference time and 77\% improvement in energy efficiency over existing state-of-the-art implementations%
    , achieving handwritten digit recognition on the MNIST dataset~\cite{MNIST} at 96.9\% accuracy at 2260 FPS, using only 0.7mJ per recognised digit.
\end{itemize}
}

\textcolor{black}{The rest of the paper is organised as follows: Section~\ref{sec:bk} covers background and related work. Section~\ref{sec:training} describes the training method used for the CNN, taking into account hardware properties. Section~\ref{sec:inference} explains further improvements for inference. Sections~\ref{sec:exp} outlines experimental results and comparisons, and finally Section~\ref{sec:conc} presents the conclusions and future directions. 
}

\section{Background and related work}\label{sec:bk}

\begin{figure*}[ht]
\begin{centering}
\includegraphics[width=.97\textwidth]{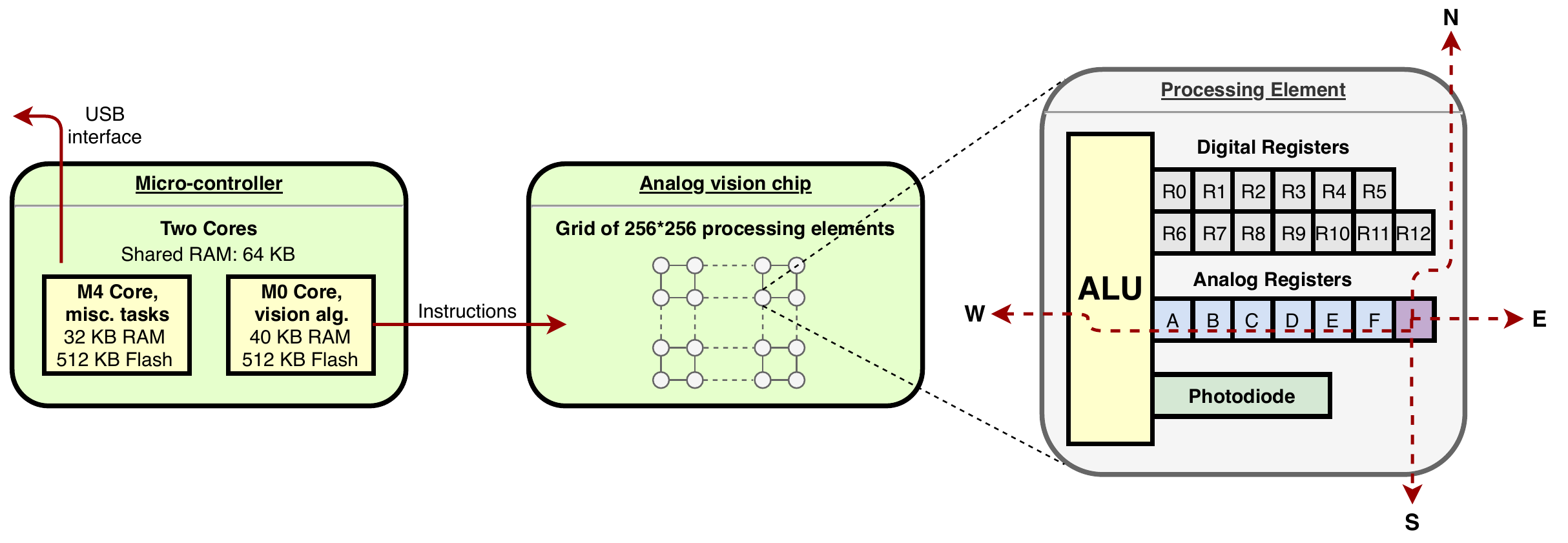}
\par\end{centering}
\begin{centering}
\caption{\label{fig:scamp5_architecture}\label{fig:scamp5_PE_registers}SCAMP5 vision system architecture: A digital micro-controller broadcasts instructions to an array of analog processing elements (PE). Each PE has an Arithmetic Logic Unit (ALU), seven local registers, and data links to its 4 neighbours.}
\par\end{centering}
\end{figure*}

\subsection{Analog Computation}
In digital computing, multiple distinct binary signals (bits) are used to represent a state or a number. In contrast, in current-mode analog computing, electrical charge is used to store a value, which can be read as an electrical current. 
The hardware needed to do arithmetic computation on analog signals is considerably simpler than on digital systems. For instance, to add two analog values, currents are joined from two sources representing the original values, while when computing in digital form using two 8-bit numbers, an adder needs many transistors (between 6 and 28 transistors per bit depending on circuit design \cite{Wairya:2012:PAH:2215602.2336658}).

\subsection{Analog Focal Plane Sensor Processors}
Focal-Plane Sensor-Processor (FPSP) chips are a special class of imaging device in which the sensor array and processor array are embedded together on the same integrated circuit \cite{Zarandy_Book_2011}. Unlike traditional vision systems, in which sensor arrays send collected data to a separate processor for processing, FPSPs allow data to be processed in place on the imaging device itself. This architecture enables ultra-fast image processing even on small, low-power devices, because costly transfers of large amounts of data are no longer necessary. Some examples of FPSPs include the ACE16k \cite{linan2002} and the MIPA4k \cite{poikonen2009}, as well as the SCAMP-5 system \cite{6578654} used to carry out the experiments described in this paper.

\begin{figure}[ht]
\begin{centering}
\includegraphics[width=.42\textwidth]{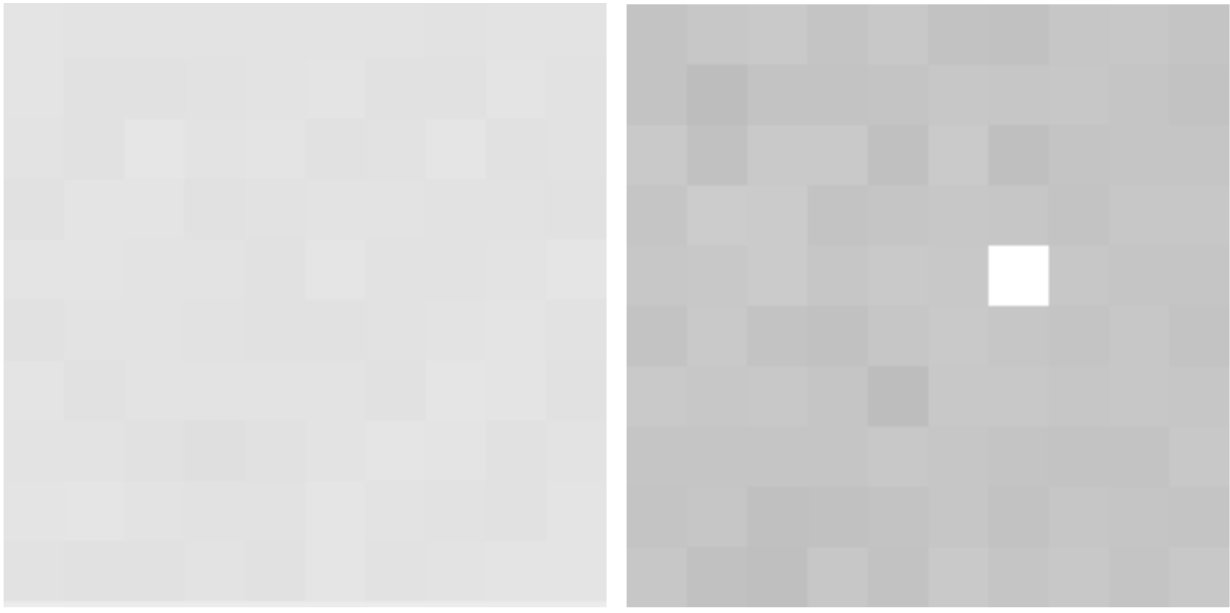}
\par\end{centering}
\begin{centering}
\caption{\label{fig:noise_demo}A real example of noisy computation on the focal plane of SCAMP5. \textbf{Left}: crop of size 10$\times$10 of a supposedly uniform input array, with value 100. \textbf{Right}: result after applying 4 times division by 2 and multiplication by 2 on this input. We notice a global shift (right image is 30\% darker), and an increased amount of noise (left image standard deviation is approx. 2, whereas it is 6.3 for the right image, with one pixel being completely off, due to computation noise).}
\par\end{centering}
\end{figure}

The SCAMP-5 device is made of a traditional \textcolor{black}{ARM Cortex M4~+ M0 dual core} digital micro-controller, connected to a light sensitive analog processor chip (the vision chip). The M4 core is responsible for sending instructions to the analog chip, while the M0 core is in charge of  miscellaneous tasks such as interfacing. The vision chip consists of a 256$\times$256 grid of pixel-processors, or Processing Elements (PEs). Arithmetic and logic instructions sent by the digital micro-controller are executed in parallel by all PEs on the array (see Figure~\ref{fig:scamp5_architecture}), on the data they locally hold. The computations are massively parallel, at the pixel level: an FPSP is a Single Instruction, Multiple Data (SIMD) computer. Additionally, the PE array is also provided with an interconnection network, allowing each PE to access values stored in its 4 neighbouring PEs' registers (see Figure~\ref{fig:scamp5_PE_registers}). 

Each PE has a light sensitive diode, so the array functions as an image sensor, and is enabled with an Arithmetic Logic Unit (ALU). 
Each ALU acts on its PE's memory registers -  only 13 digital registers each storing one bit, 7 analog registers each storing a voltage. 
\textcolor{black}{The} limited number of analog registers per pixel creates computational challenges in implementations. 

The analog nature of the computations enables high processing rates, but it introduces noise to the result. An example is shown in Figure~\ref{fig:noise_demo}. Thus it is essential to account for the impact of the noise.

The idea of reducing digital data transfers from the imaging device to the processing device is also exploited by event based cameras. Classical event based cameras such as the DAVIS240~\cite{DBLP:journals/jssc/BrandliBYLD14} or the Gen3 ATIS~\cite{prophesee} only record and transfer per-pixel intensity changes.
%A PE can be masked, to not execute the instruction broadcast to the vision chip at a certain clock tick. 
%On top of that, the PE array is also provided with a propagation network, allowing each PE to access values stored in its 4 neighbouring PEs' registers (see Figure~\ref{fig:scamp5_PE_registers}). 
%

\subsection{Vision Algorithms for FPSPs}
Vision algorithms serving various functions have been successfully implemented on analog FPSPs. Examples include FAST16 corner detection \cite{Chen_IROS_2017}, four degrees-of-freedom visual odometry \cite{Debrunner_Imperial_2017}, 
six degrees-of-freedom visual odometry \cite{murai2020bitvo},
depth estimation using a focus-tunable liquid lens~\cite{Martel_IEEE_2018}, and learning-based pixel exposure for HDR imaging and compressive sensing \cite{Martel_ICCP_2020}, \cite{MARTEL_TPAMI_2020}. 
%Chen {\it et al.} implemented a parallelized FAST16 corner detection algorithm on the SCAMP-5 at 2300 fps~\cite{Chen_IROS_2017}. Debrunner {\it et al.} implemented a simplified version of the Viola Jones Face Detection algorithm on the SCAMP-5~\cite{Debrunner_Imperial_2017}. Martel obtained depth information from scenes in real-time using a focus-tunable liquid lens in conjunction with the SCAMP-5~{\it et al.}~\cite{Martel_IEEE_2018}; they noted that their results would not have been possible using conventional hardware as doing so would have incurred a prohibitive communication overhead.

Bose {\it et al.} implemented a CNN for digit classification~\cite{bose2019camera}, \cite{bose2020fully}. Their work differs from the result of this paper in that their computations are run on digital registers, using a special ternary weight framework for convolutions - whereas we use analog registers. This achieves high speed and energy efficiency, at the price of a decrease in accuracy compared to digital implementations running on traditional hardware.

%\subsection{FPSP Code Generation}
Debrunner {\it et al.} describe AUKE~\cite{DBLP:journals/taco/DebrunnerSK19}, a code generator doing the automatic conversion of 2D convolution layers into FPSP code. Given a convolution filter, a target analog register and a set of auxiliary registers that can be used for the computations, AUKE produces the shortest program to applies the convolution on the target register. AUKE aims for the shortest possible program, for obvious execution speed reasons, but also to limit the introduction of noise to the result. \textcolor{black}{The spatial extent of the convolutional filter can be of any size, and is not restricted to immediate neighbours only (3$\times$3 convolutions)}. The limited instruction set of an FPSP however puts a restriction on the filter weights of the feasible convolutions. \textcolor{black}{The work proposed here utilizes the AUKE framework to convert convolution kernels to FPSP executable codes.}

\subsection{Hardware Acceleration of Convolutional Neural Networks}
CNN computation involves computationally demanding matrix multiplications and occurs in two modes: training and inference. While slow training may not limit their use in real-world applications, slow inference immediately becomes a bottleneck. Several attempts have been made to design customised chips capable of performing efficient convolutions~\cite{DaDianNao}, \cite{ISAAC}, \cite{ASPVision}, and \cite{Nurvitadhi_2017_FPGA}.

For instance, Origami is a power efficient and high GOp/s/W (803 GOp/s/W) digital system, devised to compute on-chip neural networks~\cite{Origami2017ITCSVT}. Chakradhar et al. designed a hardware architecture for CNNs that dynamically optimizes for performance by configuring the hardware on-the-fly~\cite{Chakradhar2010ISCA}. Similarly, Intel's Movidius Myriad~2 Neural Compute Stick (NCS), is composed of multiple cores configurable for different networks~\cite{MovidiusMyriad2}. Eyeriss~\cite{Eyeriss_IEEE_2016} is an energy efficient CNN accelerator chip, relying on data gating and compression. It runs the convolutional part of AlexNet~\cite{Krizhevsky_NIPS_2012} at 35 FPS with 278 mW power consumption.

In these designs, sensors and processor(s) are not co-located unlike in FPSPs. Instead, the hardware is configurable to run different networks.

\section{{Proposed Method: AnalogNet}} \label{sec:training}
We present a method for training a CNN architecture for image classification, that is suitable for deployment on analog FPSPs, given the availability of a training dataset $X$. We suppose $X=\left \{ (I_i,l_i) \right \}_{i=0}^N$, where each $I_i \in \mathbb{R}^{h*w*3}$ is an image, and $l_i \in \left \{ 1, ..., L \right \}$ its corresponding label. \textcolor{black}{CNN architectures for image classification start with convolutional layers, followed by fully connected layers.}

\begin{figure}[t]
\begin{centering}
\includegraphics[width=.47\textwidth]{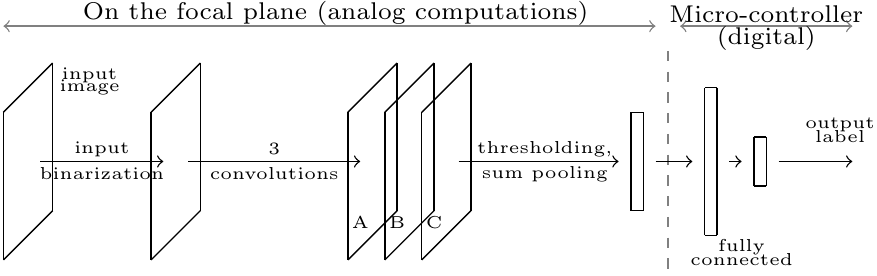}
\par\end{centering}
\caption{\label{fig:AnalogNet-Architecture}AnalogNet Architecture: convolutions are executed on the focal plane, and fully connected layers on the digital micro-controller. To reduce data transfers, only a short flat vector of scalar values is sent from the focal plane to the micro-controller, and not entire 2D feature maps.}
\end{figure}

Our method is meant for use with analog FPSPs supported by a digital micro-controller. We propose rapidly carrying out convolutions in parallel on the FPSP, followed by the transmission of intermediate results to the digital micro-controller where fully-connected layers can be computed. This architecture, making use of convolutional layers followed by fully connected layers, is a standard method in image classifiers. Instead of slowly transferring each frame to a sequential processor, one can transfer sparse feature maps which are the result of a pixel-parallel computation on this frame. Sending only short vectors of scalar (corresponding to a traditionally flattened and pooled version of 2D feature maps) greatly reduces the data transfers to the micro-controller. This is what we intend to take advantage of, as visualised on Figure~\ref{fig:AnalogNet-Architecture}. We dub the resulting architecture AnalogNet.

The training method consists in the following steps:

\begin{itemize}
\item {\bf Value Approximation and Custom
Regularization}: A standard CNN is first trained on $X$. \textcolor{black}{To do this, numerical approximation, with a custom regularization are applied, to ensure the resulting convolutions are implementable on the FPSP}.
%\item Trained weights in the convolutional layer(s) are frozen; the convolutional layer(s) are then converted into FPSP code.
\item {\bf Noise-Inclusive Training}: 
Trained weights in the convolutional layer(s) from the previous step are then converted into FPSP code using the AUKE framework. 
At this stage, the network consists of FPSP code computing the convolutional layers, followed by the fully-connected layers. A noise function is applied to each FPSP instruction in the network, and the fully connected layers of the network are re-trained on \textcolor{black}{this noisy signal}.
\end{itemize}
\textcolor{black}{Each of these steps are explained in detail below.} The result is a CNN architecture that can be used to perform inference on an analog FPSP with minimal performance loss even in the presence of hardware noise.

\subsection{Value Approximation and Custom Regularization}

Neural networks are normally implemented via multiplication of floating point numbers. However, analog FPSPs such as SCAMP-5 do not support multiplication. Instead, integer multiplication must be carried out through repeated additions.

In such cases, we resort to value approximation for multiplication. As is the case for most FPSPs, the SCAMP-5 device only supports division by two and addition. By combining repeated additions and  divisions, we can approximate the result of any floating-point multiplication operation to arbitrary precision ~\cite{DBLP:journals/taco/DebrunnerSK19}. For instance, multiplication of value $a$ by $ 0.87$ can be approximated by the following sequence:
\begin{displaymath}
0.87 \times a \approx (\frac{1}{2}+\frac{1}{4}+\frac{1}{8})\times a = 0.785 \times a
\end{displaymath}

While theoretically allowing arbitrarily precise multiplication,  in practice greater precision requires a greater number of division operations, with each operation introducing additional noise. We therefore limit approximation to a depth $n$, indicating the maximum number of divisions allowed. In practice, we used $n=2$.

We can train a neural network that accommodates these limitations, by using a regularization function that incentivises the selection of weights close to intervals of $\frac{1}{2^n}$, as such weights can be approximated with minimal error:

\begin{equation}
	L_{reg}(\theta) = cos\left(\pi(2^{n+1}\cdot\theta + 1)\right) + 1
\end{equation}

For instance, by setting $n=2$, we have $L_{reg}(\theta) = 0$ where $\theta \in \{0, \pm0.25,\pm0.5, .. \}$. We found that the most effective way of utilising the regularizer is to employ it in a \textit{re-training} loop, carried out on a network first trained without a regularizer. Weights learnt were observed to be close to the target weights, whereupon a final rounding-off was carried out.

\subsection{Noise-Inclusive Training}
At this stage convolutional layers are frozen and converted into FPSP code using AUKE \cite{DBLP:journals/taco/DebrunnerSK19}. We re-train the fully-connected layers to account for the noisy execution of the convolutional layers on the focal plane. We now introduce our approach by which hardware noise can be integrated into the training loop.

A pre-defined noise model may not always be available (as was the case for this study), as formulating a noise model is a complex task. Instead we can infer an implicit noise model directly from empirical data. The converted FPSP code obtained above is loaded onto the FPSP device, which is itself placed in front of a computer screen. Convolutions are executed on the focal plane while the dataset images are shown in sequence to the FPSP device. For each image $I_i$, we collect the flattened output of the convolutional layers $f_i \in \mathbb{R}^s$, and associate it to its ground truth label $l_i$. This process is used to create a new dataset $X'=\left \{ (f_i,l_i) \right \}_{i=0}^N$, corresponding to intermediate results of the CNN inclusive of true hardware noise. $X'$ is used to train the fully-connected layers, which implicitly learn a model of the hardware noise and compensate for it.

\section{Inference optimisation}
\label{sec:inference}
To take advantage of the pixel-parallel computations on the focal plane, we presented above how convolutions can be trained for implementation in an analog manner. Their execution on the focal plane is noisy, and their effect is only approximating what they were designed for. Conversely, the fully connected layers do not benefit from pixel parallel computations. Moreover, we expect them to compensate for the inaccurate nature of the convolutions that precede them in the network. For this reason, they are executed on the digital micro-controller which is adjacent to the focal plane.

In this section, we explain further how the result of convolutions are aggregated and sent to the micro-controller, and how the fully connected layers are executed.

\subsection{Output Event Binning Process}
The feature maps created by the convolutional part of the network are binarized: on the focal plane, locations above a certain threshold are assigned a 1, and others a 0. The corresponding threshold value is learned as part of the training process, and can be thought of as a bias for each convolution in the final layer.
The binary feature maps are then collected by the adjacent digital micro-controller in the form of a collection of \textit{events}: each event is a tuple of 2D coordinates, corresponding to the location of an activated feature. Transferring the coordinates of activated features instead of whole arrays is an effective way of reducing data transfers from the analog vision chip to the digital micro-controller.

The micro-controller is responsible for spatially binning the received events, in order to transform each feature map into a vector of fixed length. Each event falls within a spatial bin, and increments its count. This step can be seen a sum-pooling a binary feature map. As shown on Figure~\ref{fig:binning}, the binning procedure we propose uses 12 overlapping bins. Each convolutional feature map thus yields 12 values for the fully connected. This design uses the prior knowledge that the most informative locations of an image usually are in its central area. For this reason, we suggest to discard events located in the four corners, and to use different bins for the central events - to preserve more information about their location. %\textcolor{red}{Can we say anything about attention network here I though what you do is an attention mechanism towards the center of the image. Not sure though.}\textcolor{green}{Not really inspired by anything of that sort, no... It is purely guided by intuition to put the bins here.}

\definecolor{eb1}{HTML}{990000}
\definecolor{eb2}{HTML}{994C00}
\definecolor{eb3}{HTML}{999900}
\definecolor{eb4}{HTML}{4D9900}
\definecolor{eb5}{HTML}{009900}
\definecolor{eb6}{HTML}{00994D}
\definecolor{eb7}{HTML}{009999}
\definecolor{eb8}{HTML}{004C99}
\definecolor{eb9}{HTML}{000099}
\definecolor{eb10}{HTML}{4C0099}
\definecolor{eb11}{HTML}{990099}
\definecolor{eb12}{HTML}{99004D}

\begin{figure}[t]
\begin{centering}
\includegraphics[width=.25\textwidth]{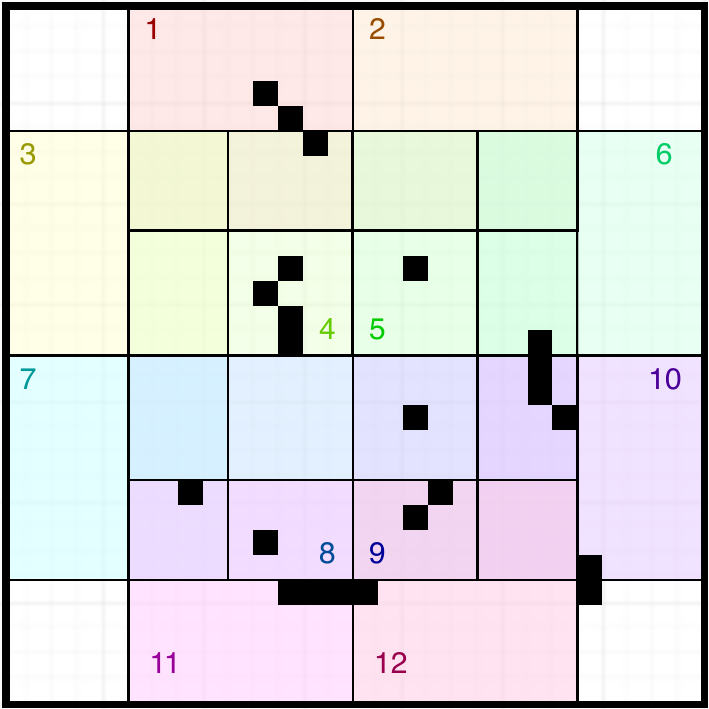}
\par\end{centering}
\begin{centering}
\caption{\label{fig:binning}AnalogNet's event binning process. Events (i.e. activated binary features) fall into twelve overlapping bins. Black pixels correspond to a possible events configuration, here yielding the output vector $\left ( {\color{eb1} 3}, {\color{eb2} 0}, {\color{eb3} 0}, {\color{eb4} 5}, {\color{eb5} 2}, {\color{eb6} 1}, {\color{eb7} 1}, {\color{eb8} 2}, {\color{eb9} 6}, {\color{eb10} 4}, {\color{eb11} 5}, {\color{eb12} 3} \right )$.}
\par\end{centering}
\end{figure}

\subsection{Fully Connected Layers Computations}

The vectors resulting from the above binning process (one per feature map) are aggregated into a single vector, and given as input to a fully connected network. Classically, taking the argmax of the final layer's output produces the predicted label for the input image.

As introduced earlier, the FPSP's micro-controller is very rudimentary, and clocked at low frequency. Digital computations can thus easily become the speed bottleneck of an FPSP program. Consequently, to not hinder the great advantage brought by the execution of convolutions on the focal plane, the digital computations of the fully connected layers had to be optimised. The execution of a single fully connected layer can be seen as matrix-vector multiplication, with the addition of a point-wise bias - the matrix carries the weight of the layer, and the vector consists in its input values. A standard implementation relies on double loops for matrix-vector multiplication.

In our case, once the training is done, the network structure and weights are fixed. As a result, we know in advance (at compilation time), the number of iterations needed in each loop, and the coefficient of each multiplication. For this reason, we unroll the loop before compilation, and hard-code the weights in the unrolled loop. This removes the overhead of accessing memory, and even frees some RAM. The length of the program is considerably increased, but the resulting speed-up is substantial, as presented in Table~\ref{tab:fc_loop_unrolling_time_comparison}. This technique is equivalent to putting some manual optimisation upstream of the compiler.

\begin{table}[t]
\caption{\label{tab:fc_loop_unrolling_time_comparison}Comparison of computation time of a 2 layers fully connected network (36 input units, 50 hidden, 10 output), with three different methods used for matrix-vector multiplication, on the SCAMP-5 vision system.}
\begin{center}
\begin{tabular}{lc@{\quad}cl}%{l@{\quad}r@{\quad}rl}
\hline
\multicolumn{1}{l}{\rule{0pt}{12pt}
                   Method}&
                   \begin{tabular}[c]{@{}c@{}}{Time for fully-connected}\\ {computation ($\mu$s)}\end{tabular}&
                   &\\[2pt]
\hline\rule{0pt}{12pt}
Standard double loop   &   483 &\\\rule{0pt}{0pt}
Unrolled loop               &   465 &\\\rule{0pt}{0pt}
\begin{tabular}[|c|]{@{}l@{}}\textbf{Ours}: unrolled loop \\\quad and hardcoded weights\end{tabular}        &   \textbf{136}&\\
[2pt]
\hline
\end{tabular}
\end{center}
\end{table}

\section{Experiments}
\label{sec:exp}
We conducted experiments demonstrating that a network architecture trained and implemented using the method described above can successfully achieve competitive performance on FPSP hardware. We then evaluated it on the MNIST hand-written digit classification dataset~\cite{MNIST}, comparing the accuracy, inference time and power consumption of this architecture as implemented on FPSP (SCAMP-5), CPU (Intel i7-4930K), GPU (Nvidia GTX1080) and Vision Processing Unit (VPU, Intel Myriad2 Neural Compute Stick).

\subsection{Architecture}
%\textcolor{red}{This paragraph need more edit, there was nothing about 2-layer previously. I think you intentionally did not include details about 2-layer networks. One idea would be that you could at least briefly explain one of 2-layer architecture, the one that seems more reasonable, explain it and compare it with 1- layer. And then express that because of noise problems with deeper networks, we focus on a-layer only. What do you think? }\textcolor{green}{ I agree!}

\textcolor{black}{Here we explain the reasoning and experiments resulting in the final choice of architecture for AnalogNet.}

\subsubsection{Single convolutional layer}

We first showed the relevance of our training method in the case of a single layer CNN. In this single layer, the number of convolutions remains to be chosen. For this reason, we studied the accuracy of single layer CNNs using increasingly many convolution kernels (or \textit{feature maps}). All other parameters are fixed: each convolution kernel is of size 3$\times$3, each feature map is pooled according to the binning method presented above, and the fully connected layer has two layers - 50 hidden units, 10 output units.

\begin{figure}[t]
\begin{centering}
\includegraphics[width=.38\textwidth]{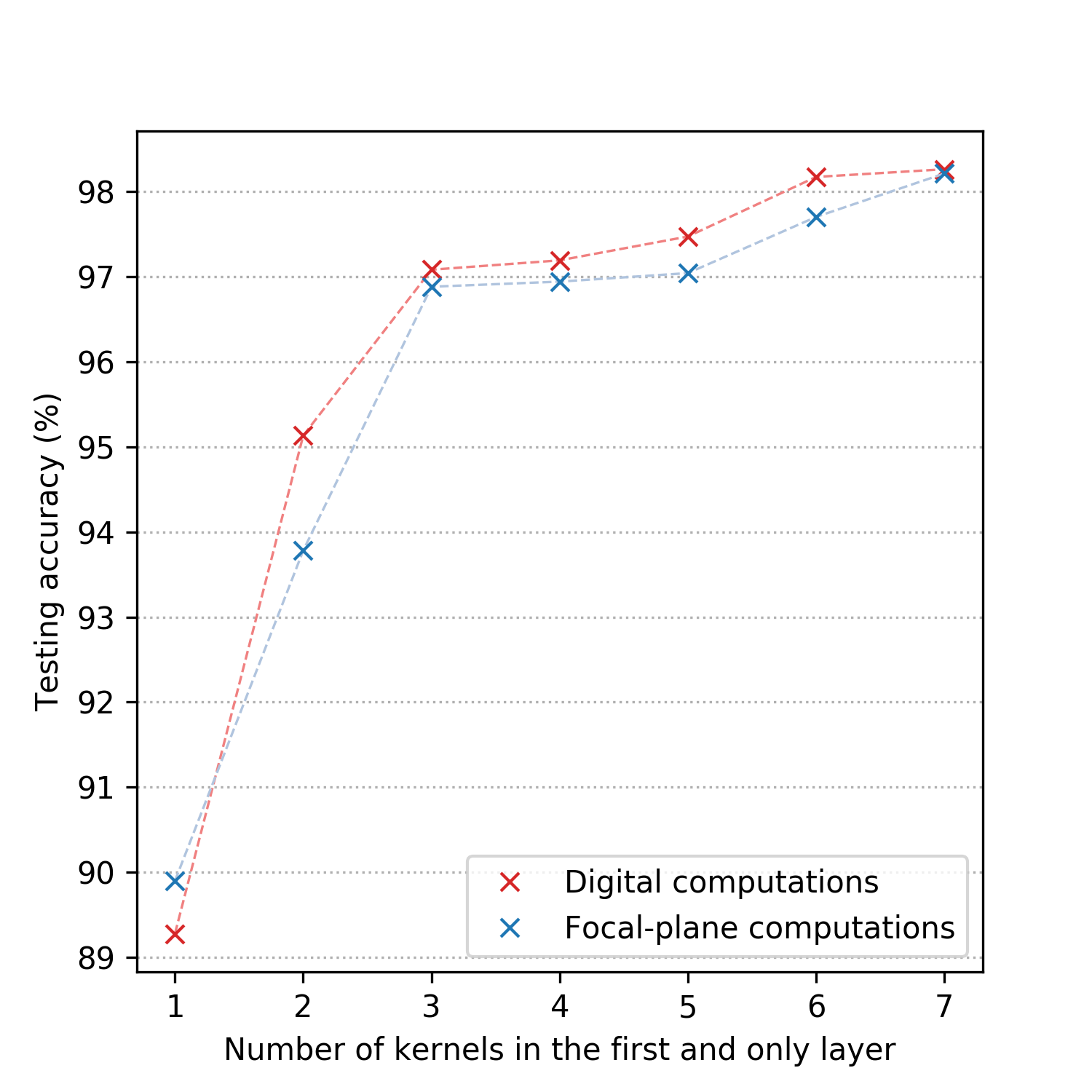}
\par\end{centering}
\caption{\label{fig:increase_kernels}Testing accuracy, with increasingly many kernels in a single layer CNN.}
\end{figure}

\begin{figure}[t]
\begin{centering}
\includegraphics[width=.38\textwidth]{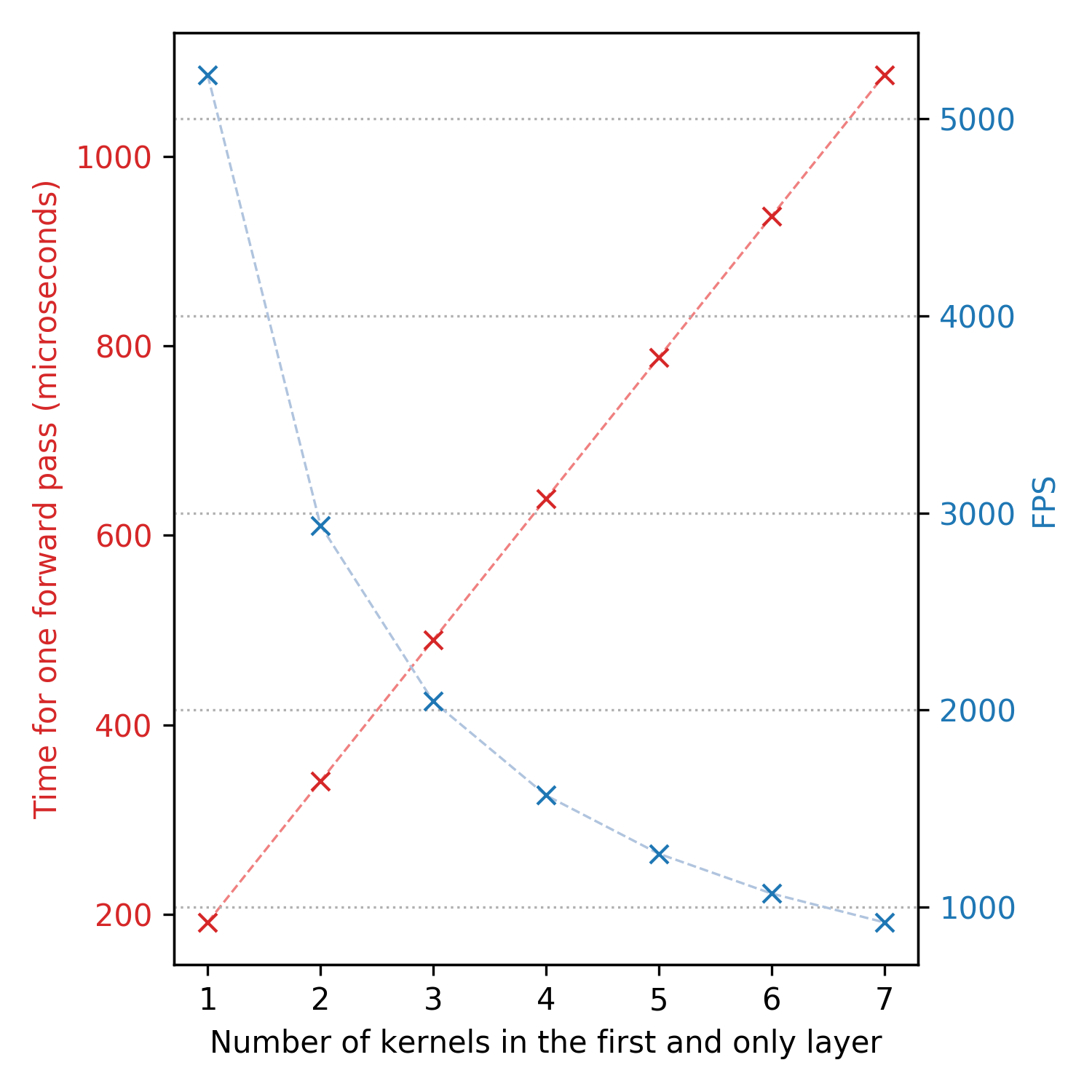}
\par\end{centering}
\caption{\label{fig:increase_kernels_time}Estimation of the time required for one forward pass of a single layer network with increasingly many kernels, and the resulting FPS.}
\end{figure}

Figure~\ref{fig:increase_kernels} reports the resulting testing accuracy, both for a digital implementation in Tensorflow~\cite{abadi2016tensorflow}, and for a final implementation on the SCAMP5 device. For each kernel number between $1$ and $7$, the full training process is run. For a given kernel number, the testing accuracy reported for the digital implementation is greater than the final implementation on the SCAMP-5 device: this is due to noisy nature of computations on the focal plane. As one could expect, a network with more kernels reaches a higher accuracy, but the returns of adding new kernels is diminishing. The drawbacks of using more kernels are an increased latency and power consumption of the convolutional part, which scale linearly with the number of feature maps it has. Figure~\ref{fig:increase_kernels_time} presents estimated inference duration and frames per second.\\

This systematic search of the architecture space of single layer CNNs clearly exhibits the trade-off at stake. We consider the three kernels, single layer CNN to be at the sweet spot of this design space. With more kernels, the networks fall below the landmark of 2000 FPS. Being at the inflexion point on Figure~\ref{fig:increase_kernels}, adding more than 3 kernels gives diminishing returns in terms of testing accuracy.

\subsubsection{Experimenting with two layers of convolutions}

We then briefly explored the feasibility of implementing multiple convolutional layers on the SCAMP-5 device. The main design issue that arises is caused by the very limited register availability. In a traditional 2-layer CNN, each feature map of the second convolutional layer is the sum of the results of convolutions applied to each feature map in the first layer. This creates the need to store partial results, since all feature maps of the first layer are required until the very last feature map of the second layer is computed. \textcolor{black}{This is a challenging problem, due to limited number of analog registers (AREGs) available for each pixel.}

\textcolor{black}{To store more partial intermediate feature maps, we tried storing some analog values in boolean digital registers (DREGs) for later use, when not presently needed for any computation. Using a quantisation procedure, one AREG's content can be encoded in binary and moved to multiple DREGs. The AREG can then be freed and used for current computations, while its content is saved and available for later use. An AREG stores a voltage representing integer values between -127 and 127, while a DREG stores either a 0 or a 1.}
\textcolor{black}{These A/D conversions are extremely fast, and data remains on the focal plane - we do not give up to the benefits of FPSP computations. The precision level of the quantisation procedure can be adjusted, to choose between very frugal DREG use and high precision quantisation. Theoretically, the whole dynamic range of an AREG requires 8 DREGs to be precisely stored. In practice, operations are noisy, and using 8 DREGs provides an unnecessary amount of precision. In our experiments, we found 3 to 4 DREGs to be sufficient to capture the content of one AREG accurately enough.}

\textcolor{black}{Quantization is an idea that has already been explored to reduce the memory footprint of CNNs, as in \cite{DBLP:journals/corr/HanMD15}, \cite{DBLP:journals/corr/ZhouYGXC17} or \cite{DBLP:journals/corr/CourbariauxB16}. In our case, we used a two layer CNN, each layer having 3 feature maps. The quantisation procedure is used to binarise the first layer's feature maps when required. The best result we could achieve showed a final accuracy more than 1\% below the single layer one that uses 3 kernels. We therefore decided not to use it as our final implementation of AnalogNet.}

\textcolor{black}{We also experimented another technique based on spatial pooling. Pooling feature maps decreases their spatial dimensionality, allowing for multiple ones to be stored on a single AREG. This idea of interleaving multiple arrays of data on a single AREG was explored by \cite{DBLP:conf/ecctd/MartelC0D15}, but  showed disappointing performance in our case.}

\textcolor{black}{Our hypothesis is that the poor performance of two-layer convolutional layers on the SCAMP-5 device is mostly due to noise accumulating at each operation. For this reason, the final AnalogNet architecture consists of a single convolutional layer, with 3 convolutions.}

\subsubsection{Noise accumulation}

%To demonstrate the negative effect of noise accumulation without relying on an unusual and complex ways of running multiple layers on the SCAMP-5 device, we assess
To demonstrate the negative effect of noise accumulation when running multiple layers of convolutions on the SCAMP-5 device, the performance of \textit{depth-separable} CNNs is assessed. In depth-separable convolutions, feature maps are not combined with summation as with traditional convolutional layers. Instead, series of convolutions and non-linearities are applied to the input images, each one forming an independent thread in the computational graph. Depth-separable convolutions are most often used to reduce memory footprint of CNNs \cite{MobileNet_arxiv_2017}. In our case, this greatly decreases the requirement to store partial results on the focal plane. Since feature maps are not combined, feature extraction is not as powerful and efficient as with standard convolutional layers. However, the accumulation of non-linearities and convolutions still helps in improving testing accuracy.

\begin{figure}[t]
\begin{centering}
\includegraphics[width=.38\textwidth]{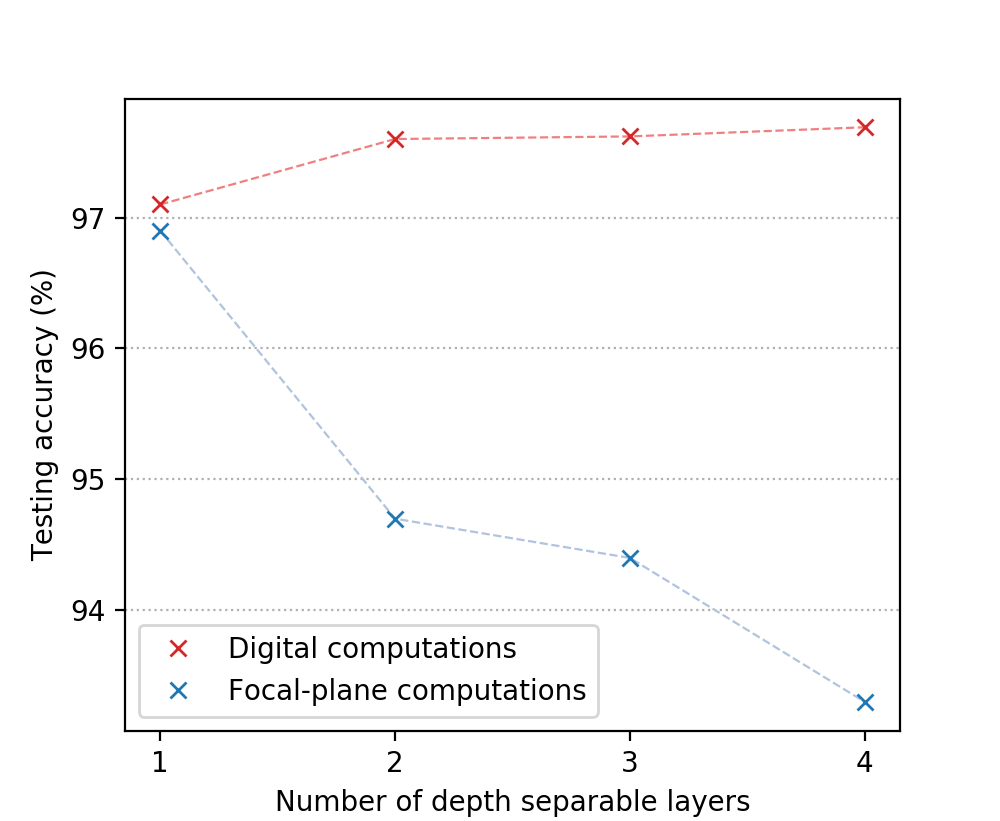}
\par\end{centering}
\begin{centering}
\caption{\label{fig:depth_sep_conv_accuracies}MNIST testing accuracy of CNNs using increasingly many layers of depth separable convolutions. Each layer is made of three 3*3 convolutions, and a ReLU activation function. The case with 1 layer corresponds to AnalogNet.}
\par\end{centering}
\end{figure}

Figure~\ref{fig:depth_sep_conv_accuracies} shows the MNIST testing accuracy of CNNs using increasingly many layers of depth separable convolutions, both in our noiseless digital simulations and on the SCAMP5 device. In simulations, having more depth-separable layers yields better results, with diminishing returns. On the SCAMP5 device, the exact inverse phenomenon happens, with results getting worse as each new layer is added.

This results confirm our hypothesis that despite being more powerful in theory - i.e. with noiseless and precise digital implementations - networks involving longer chains of computation perform poorly on the SCAMP5 device. The computation chain between the input image and the output feature maps is longer, and the information is supposed to be refined, but is in reality diluted in increasing amounts of noise. What really is detrimental here is the accumulation of noise.

\subsection{Accuracy}
In Table~\ref{tab:cnns_accuracy}, we compare AnalogNet's accuracy on the MNIST test dataset with the accuracy reported by Bose et al.~\cite{bose2019camera}. \textcolor{black}{Similarly to the noise aware training process presented in Section~\ref{sec:training}, the FPSP is placed in front of a computer screen, which is displaying the whole MNIST test set. The label predicted by the SCAMP-5 is compared to the ground truth one.}

Our solution achieves decent \textcolor{black}{accuracy, at 96.9\%}. However, although this aspect is not \textcolor{black}{quantified}, the method from Bose et al. seems more robust to image misalignment \textcolor{black}{and variable point of view}. We also compared to a digital (noiseless) implementation of AnalogNet, and RMDL~\cite{kowsari2018rmdl}, which achieves state of the art MNIST classification accuracy, but relies on a much more complex architecture than AnalogNet.

\begin{table}[t]
\caption{\label{tab:cnns_accuracy}MNIST testing accuracy comparison. A \textit{digital} device refers to either CPU, GPU or VPU.}
\vspace{-4 mm}
\begin{center}
\begin{tabular}{l@{\quad}@{\quad}@{\quad}ccl}
\hline
\multicolumn{1}{l}{\rule{0pt}{8pt}
                   Architecture}&
                   {Hardware}&
                   {Testing accuracy}&\\[0pt]
\hline\rule{0pt}{4pt}
AnalogNet              & SCAMP-5     &   \textbf{96.9\%} &\\\rule{0pt}{0pt}
Bose~\cite{bose2019camera} & SCAMP-5 &   94.2\% &\\\rule{0pt}{0pt}
AnalogNet & digital &   97.1\% &\\\rule{0pt}{0pt}
Bose~\cite{bose2019camera} & digital &   95.4\% &\\\rule{0pt}{0pt}
\textit{RMDL~\cite{kowsari2018rmdl}}    & \textit{digital} &   \textit{99.82\%} &\\
[2pt]
\hline
\end{tabular}
\vspace{-1.5 mm}
\end{center}
\end{table}

\subsection{Inference Time}
\iffalse
\begin{table}[t]
\caption{\label{tab:inference_time}Estimated per-frame computation time in micro-seconds, for AnalogNet architecture. Image data retrieval time is conservatively lower bounded for CPU, GPU, and VPU, while it effectively is zero for SCAMP-5.}
\vspace{-4 mm}
\begin{center}
\begin{tabular}{c@{\quad}@{\quad}cccl}
\hline
\multicolumn{1}{l}{\rule{0pt}{8pt}
                   Hardware}&
                   {Image retrieval}&
                   {Inference}&
                   {Total}&\\[0pt]
\hline\rule{0pt}{12pt}
SCAMP-5 & 0          & 442 & \textbf{442}\\\rule{0pt}{0pt}
VPU     & $\geq$ 100 & 420 & $\geq$ 522\\\rule{0pt}{0pt}
GPU     & $\geq$ 100 & 758 & $\geq$ 858\\\rule{0pt}{0pt}
CPU     & $\geq$ 100 & 651 & $\geq$ 751\\
[2pt]
\hline
\end{tabular}
\vspace{-1.5 mm}
\end{center}
\end{table}
\fi

\begin{table}[t]
\caption{\label{tab:inference_time}Estimated per-frame computation time in micro-seconds, for AnalogNet architecture. Image data retrieval time is excluded for CPU, GPU, and VPU, while it effectively is zero for SCAMP-5.}
\vspace{-4 mm}
\begin{center}
\begin{tabular}{c@{\quad}@{\quad}cl}
\hline
\multicolumn{1}{l}{\rule{0pt}{8pt}
                   Hardware}&
                   {Inference time}&\\[0pt]
\hline\rule{0pt}{12pt}
SCAMP-5 & 442\\\rule{0pt}{0pt}
VPU     & 422\\\rule{0pt}{0pt}
GPU     & 758\\\rule{0pt}{0pt}
CPU     & 651\\
[2pt]
\hline
\end{tabular}
\vspace{-1.5 mm}
\end{center}
\end{table}

For the SCAMP-5, VPU, CPU and GPU, inference time was estimated by measuring the average per-frame computation time incurred on each system. It was measured by their respective system utilities. Table~\ref{tab:inference_time} shows the recorded per-frame computation time. Excluding the substantial time cost of capturing and retrieving the image data from a separate camera for computation on the CPU, GPU and VPU, the SCAMP-5 FPSP is already much faster than the CPU and the GPU, and reaches almost the same speed as the VPU. \textcolor{black}{Note that these figures include the total time for inference, i.e. both the convolutional part and the fully connected one - and the transfer of binary features from the focal plane to the micro-controller in the case of SCAMP-5.}

In practice this data retrieval bottleneck is a major limitation on frame rates achievable on CPU, GPU and VPU. The FPSP does not face this bottleneck because image data is processed in place at the pixel level. \iffalse We lower bound the image capturing and retrieval time by 100 $\mu$s. This corresponds to the average time it takes to load a single MNIST digit image from the SSD of a modern laptop to its RAM. This value is extremely conservative compared to actually transferring an image from a camera to a computation device. It can safely be considered as a lower bound to a situation dependent value. %\textcolor{red}{TODO: Maybe you could give an example for upper bound, just to give the reviewer an idea.} 
\textcolor{black}{To put this figure into perspective, transferring an 8 bit image of size 28$\times$28 pixels from the sensor of the SCAMP-5 camera (which can to some degree also be used as a standard webcam) to a nearby computer via USB 2.0 takes 1584 $\mu$s.}\fi
Depending on the application, the image retrieval process can be fully parallelized, and be faster than the inference time. In these cases, in the continuous regime, the data retrieval bottleneck vanishes. In other situations, a real time, low latency prediction is needed - such as in high speed robotics. In these cases, the latency of a standard camera can become the first obstacle in achieving low latency. For instance, transferring an 8 bit image of size 28$\times$28 pixels from the sensor of the SCAMP-5 camera (which can also be used as a standard webcam) to a nearby computer via USB 2.0 takes 1584 $\mu$s.

Taking this into account shows the clear advantage of realising most computations directly on the light sensor, in an analog manner. The image retrieval latency is simply reduced to zero. \textcolor{black}{Implemented on the SCAMP-5 device, AnalogNet infers the label of an MNIST digit in 442 $\mu$s, giving an framerate of 2260 FPS. The latter is a calculated figure, since our computer display setup cannot achieve this refresh rate.}

AnalogNet compares very favourably to Bose~\cite{bose2019camera}, which reports a maximum frame-rate of 210 FPS, equivalent to a minimum inference time of approximately 4700$\mu$s. Note that the GPU inference time is disappointingly high. Our speculation is that GPUs are optimised for treating large batches of images, but not single instances sequentially, in an online manner.

\subsection{Energy Consumption}
\begin{table}[t]
\caption{\label{tab:energy_consumption}Estimated per-frame energy consumption, in milli-Joules. Energy consumption incurred in capturing and retrieving an image is excluded for CPU, GPU, and VPU, while the SCAMP-5 figure is `all-in'. For the VPU and CPU, energy drawn by the host computer is conservatively lower bounded.}
\vspace{-4 mm}
\begin{center}
\begin{tabular}{c@{\quad}@{\quad}cccl}
\hline
\multicolumn{1}{l}{\rule{0pt}{8pt}
                   Hardware}&
                   {Host}&
                   {Inference}&
                   {Total}&\\[0pt]
\hline\rule{0pt}{12pt}
SCAMP-5 & 0         & 0.7  & \textbf{0.7}\\\rule{0pt}{0pt}
VPU     & $\geq$0.9 & 0.3  & $\geq$1.2\\\rule{0pt}{0pt}
GPU     & $\geq$1.5 & 28.6 & $\geq$30.1\\\rule{0pt}{0pt}
CPU     & 0         & 18.9 & 18.9\\
[2pt]
\hline
\end{tabular}
\vspace{-1.5 mm}
\end{center}
\end{table}

Per-frame energy consumption was determined by estimating the power drawn by each device multiplied by the per-frame inference time. The power draw was measured using Nvidia SMI for the GPU, Intel RAPL for the CPU, and an USB power meter for the VPU and the SCAMP-5. Table~\ref{tab:energy_consumption} shows the estimated per frame energy consumption. AnalogNet inference on the SCAMP-5 FPSP uses far less energy per-frame than a CPU or a GPU and is comparable to a low-power VPU.

Note that the FPSP figure is an `all-in' one, while the CPU, GPU and VPU estimates do not consider the energy consumption required for image capture and data transfer, i.e. the energy cost of capturing and retrieving the image data from a separate camera or a digital storage. Furthermore, energy consumption estimates for the VPU and GPU have to account the power drawn by a host computer. This is conservatively lower bounded by 2 Watts, which is the power drawn at idle by a Raspberry Pi 3.

\subsection{Discussion: Limitations}
%\subsubsection{Limitations}

While state-of-the-art CNN architectures use many convolutional layers, our work was limited to a single convolutional layer, thereby restricting our CNN experiments to the MNIST dataset. There are two fundamental challenges involved in computing additional layers on an analog FPSP:
\begin{itemize}

\item \textbf{Data Loss:} In analog systems, every computation introduces additional noise to the original signal. Additional layers naturally require a greater number of computations.

\item \textbf{Data Sparsity:} Data is stored geographically on the FPSP, resulting in pooled data being spatially dispersed. To continue computation, the data needs to be spatially compressed, introducing additional noise. 
\end{itemize}

In this application, only the central 28$\times$28 subarray of a 256$\times$256 device is used, and hence the computational capabilities of the focal plane are not fully exploited.

% and lack of memory available at each PE. 

%\subsubsection{Further Work}
%Further work in this area should focus both on designing additional network architectures capable of operating within current hardware constraints, and on developing new FPSP hardware designs that alleviate the limitations encountered in this study. One significant hardware contribution would be more memory at each processing element (PE). Currently, SCAMP-5 PEs have access to extremely limited memory. Additional memory (analog or digital) would facilitate implementation of complex neural networks, with an increase in the amount of memory available directly corresponding to an increase in the ability of the FPSP to handle convolutional layers of greater breadth.

%Accounting for noise for CNN inference also remains an important issue. It could better be integrated in the  training, if a precise differentiable noise model was available. Another potential solution could be to mitigate the effects of noise at inference time, for instance by running each operation multiple times and averaging their result.

\section{Conclusion}
\label{sec:conc}
In this paper, we demonstrated the potential for analog FPSPs to bring computer vision capabilities embedded applications that are constrained by power budgets and requiring very low latency, such as robotics. We developed a method for designing customised CNNs for use on analog FPSPs that are able to overcome the challenges inherent in working with such devices.
We performed experiments on the MNIST dataset, showing that digit recognition using a CNN can indeed be carried out on an analog FPSP.
By using FPSPs to process data in the focal plane and limit\textcolor{black}{ing} data movement only to important features, we showed that we can increase the frame-rate and reduce the overall power consumption significantly. 

%In future work, we aim to design FPSPs with more resources in the focal plane that are able to run complex pipelines and deeper networks. This can be achieved by co-designing software and hardware, taking into account the optimal configuration for the overall pipeline.

\textcolor{black}{We realise that the greatest limitation of our work is the very simple architecture of AnalogNet in regard to the current state of deep learning.} Further work should focus both on designing additional network architectures capable of operating within current hardware constraints, and on developing new FPSP hardware designs that alleviate the limitations encountered in this \textcolor{black}{paper}. One significant hardware contribution would be more memory at each processing element (PE). Currently, SCAMP-5 PEs have access to extremely limited memory. Additional memory (analog or digital) would facilitate implementation of complex neural networks, with an increase in the amount of memory available directly corresponding to an increase in the ability of the FPSP to handle convolutional layers of greater breadth. 

Accounting for noise for CNN inference also remains an important issue. It could better be integrated in the  training, if a precise differentiable noise model was available. Another potential solution could be to mitigate the effects of noise at inference time, for instance by running each operation multiple times and averaging their result.

%% Acknowledgments section, defined using the "acks" environment
\begin{acks}
This research is supported by Engineering and Physical Sciences Research Council [grant number EP/K008730/1], A Panoramic View of the Many-core Landscape (PAMELA). We thank Piotr Dudek and his colleagues, Stephen Carey and Jianing Chen, at the University of  Manchester for kindly providing access to the SCAMP device and simulator.
\end{acks}

%% bibliography style to be used
\bibliographystyle{ACM-Reference-Format}
%% bibliography file, without the ".bib" extension
\bibliography{mybib}

%% Appendix?
%\appendix

\end{document}